\documentclass{PoS}%
\usepackage{amsmath}
\usepackage{amsfonts}
\usepackage{amssymb}
\usepackage{graphicx}%
\providecommand{\U}[1]{\protect\rule{.1in}{.1in}}

\title{%
\protect{   \vspace{-2cm} 
\flushright{
\begin{minipage}{6cm}
    \normalsize 
    KEK Preprint 2012-35 \\
    CHIBA-EP-198 
\end{minipage} 
}\\
\vspace{2cm}%
}
Gluon propagators in the deep IR region and non-Abelian dual superconductivity for SU(3) Yang-Mills}

\ShortTitle{Gluon propagators and non-Abelian dual superconductivity for SU(3) Yang-Mills }

\author{\speaker{Akihiro Shibata}\\
        Computing Research Center, High Energy Accelerator Research Organization (KEK) \&  \\
        Graduate University for Advanced Studies (Sokendai), Tsukuba 305-0801, Japan\\
        E-mail: \email{akihiro.shibata@kek.jp}}
\author{Kei-Ichi Kondo\\
        Department of Physics, Graduate School of Science, Chiba University, Chiba 263-8522, Japan\\
        E-mail: \email{kondok@faculty.chiba-u.jp}}
\author{Seikou Kato\\
        Fukui National College of Technology, Sabae, Fukui 916-8507, Japan\\
        E-mail: \email{skato@fukui-nct.ac.jp}}
\author{Toru Shinohara\\
        Department of Physics, Graduate School of Science, Chiba University, Chiba 263-8522, Japan\\
        E-mail: \email{sinohara@graduate.chiba-u.jp}}

\FullConference{The 30 International Symposium on Lattice Field Theory - Lattice 2012,\\
		June 24-29, 2012\\
		Cairns, Australia}

\abstract{%
    We have presented non-Abelian dual superconductivity picture in the SU(3) Yang-Mills(YM) theory, and shown evidences such as the restricted U(2)-field dominance and the non-Abelian magnetic monopole dominance in the string tension. To establish the dual superconductivity picture, the dual Meissner effect in Yang-Mills theory must be examined, and we also presented the evidence of non-Abelian dual Meissner effect by measuring chromo-electric flux tube in the last lattice conferences.  \\
    In this talk, by applying a new formulation of the YM theory on a lattice, the we further investigate the non-Abelian dual Meissner effect for SU(3) YM theory through correlation function. We examine non-abelian magnetic monopole currents as well as color flux created by the quark-antiquark source.}

\begin{document}
\section{Introduction}

Quark confinement follows from the area law of the Wilson loop average. The
dual super conductivity is the promising mechanism for quark confinement
\cite{dualSC}. Based on the Abelian projection, there have been many numerical
analyses such as Abelian dominance \cite{Suzuki90}, magnetic monopole
dominance \cite{stack94} \cite{shiba}, and center vortex dominance
\cite{greensite} in the string tension. However, these results are obtained
only in special gauges such as the maximal Abelian (MA) gauge and the
Laplacian Abelian gauge, and the Abelian projection itself breaks the gauge
symmetry as well as color symmetry (global symmetry).

We have presented the lattice version of a new formulation of $SU(N)$
Yang-Mills (YM) theory\cite{KSM05}\cite{SCGTKKS08}, that gives the
decomposition of the gauge link variable suited for extracting the dominant
mode for quark confinement in the gauge independent way. In the case of the
$SU(2)$ YM theory, the decomposition of the gauge link variable is given by a
compact representation of Cho-Duan-Ge-Faddeev-Niemi (CDGFN)
decomposition\cite{CFNS-C} on a lattice \cite{ref:NLCVsu2}\cite{ref:NLCVsu2-2}%
\cite{kato:lattice2009}. For the $SU(N)$ YM theory, the new formula for the
decomposition of the gauge link variable is constructed as an extension of the
$SU(2)$ case. There are several possibilities of decomposition corresponding
to the stability subgroup $\tilde{H}$ of gauge symmetry group $G,$ while there
is the unique option of $\tilde{H}=U(1)$ in the $SU(2).$ For the case of
$G=SU(3)$, there are two possibility which we call the maximal option and the
minimal option. The maximal option is obtained for the stability group
$\tilde{H}=U(1)\times U(1)$, which is the gauge invariant version of the
Abelian projection in the maximal Abelian (MA) gauge \cite{lattce2007}. The
minimal one is obtained for the stability group of $\tilde{H}=U(2)\cong
SU(2)\times U(1)$, which is suitable for the Wilson loop in the fundamental
representation derived from the non-Abelian Stokes'\ theorem \cite{KondoNAST}.
For the quark-antiquark (fundamental representation of ) static potential, we
have demonstrated the gauge independent (invariant) restricted $U(2)$%
-dominance, (or conventionally called "Abelian\textquotedblright\ dominance),
$(\sigma_{V}/\sigma_{full}=93\pm16\%)$ where the decomposed $V$-field
(restricted U(2) field) reproduced the string tension of original YM field,
and \ the gauge independent non-Abelian magnetic monopole dominance
($\sigma_{mon}/\sigma_{V}=94\pm9\%$), where the string tension was reproduced
by only the (non-Abelian) magnetic monopole part extracted from the restricted
U(2) field \cite{SCGTKKS08L}\cite{lattice2008}\cite{lattice2009}%
\cite{lattice2010}\cite{abeliandomSU(3)}

To establish the dual superconductivity picture, we must also show the
magnetic monopoles play the dominant role in quark confinement. The dual
Meissner effect in Yang-Mills theory must be examined by measuring the
distribution of chromo-electric field strength or color flux as well as
magnetic monopole currents created by a static quark-antiquark. In
$SU(2)$\ case, the extracted filed corresponding to the stability group
$\tilde{H}=U(1)$ shows the dual Meissner effect \cite{DualSC:KKSS2012}, which
is a gauge invariant version of the Abelian projection in MA gauge. In the
SU(3) case, there are many works on color flux for the Yang-Mills field by
using Wilson line/loop operator, e.g., \cite{Cardaci2011} \cite{Cardso}
\cite{CeaCosmail2012} However, there is no direct measurement of the dual
Meissner effect in the gauge independent (invariant) way, except for several
studies based on the Abelian projection, e.g., \cite{suzuki:ejiri} By applying
our new formulation to the $SU(3)$ YM theory, we have shown the evidence of
the non-Abelian dual Meissner effect claimed by us by measuring the color flux
created by a static quark-antiquark pair, and found the color flux tube.

In this talk, we further study the non-Abelian dual Meissner effect and the
correlation functions (propagator) for the original YM field and the
decomposed variables.

\section{Method}

\begin{figure}[ptb]
\begin{center}
\vspace{-7mm}\includegraphics[
height=4.5cm,
]
{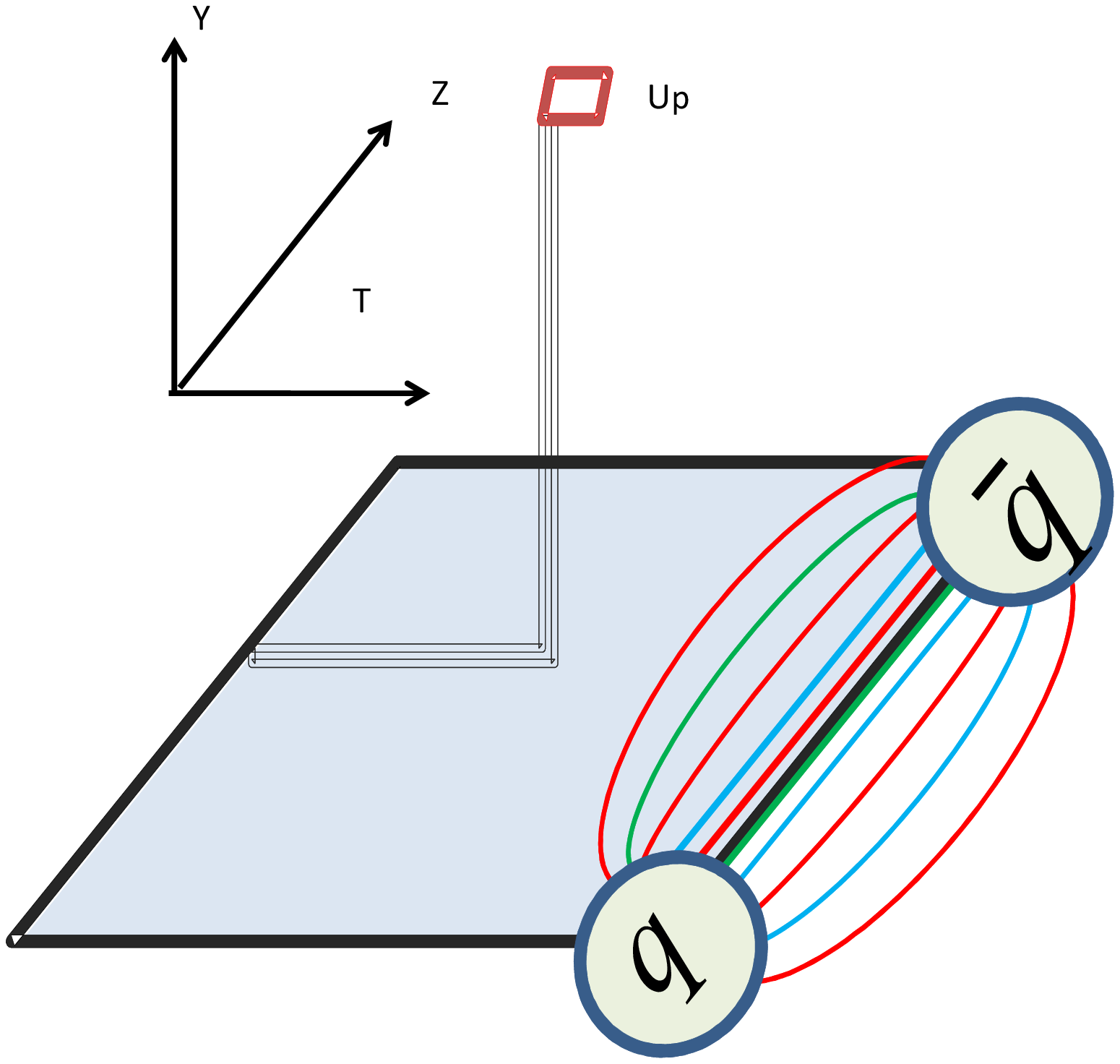} \ \begin{minipage}{0.45\textwidth}
\includegraphics[
height=7cm,
angle=270,
origin=t
]
{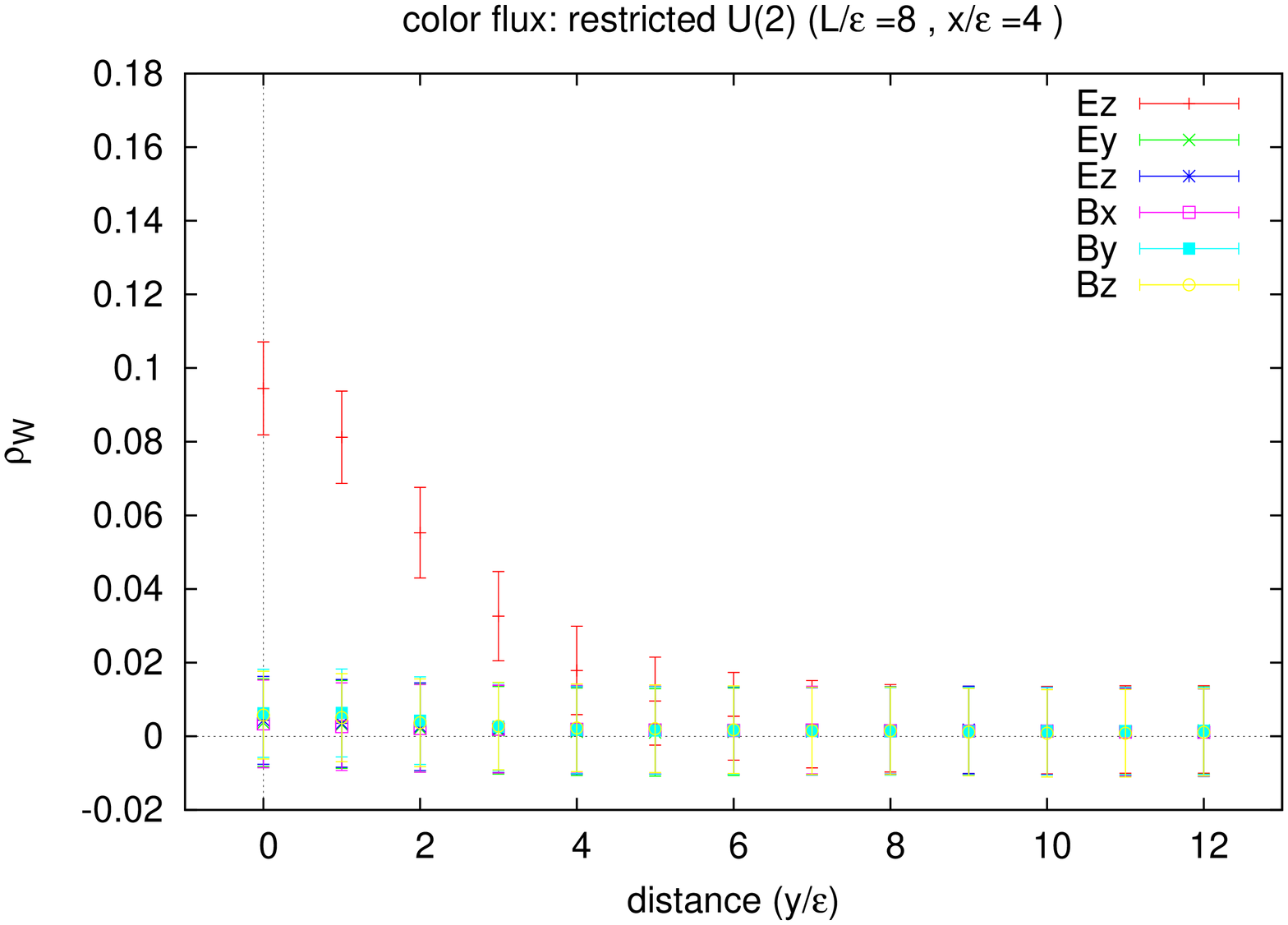}
\end{minipage}
\vspace{-5cm}
\end{center}
\caption{(Left) The connected correlator ($U_{p}LWL^{\dag})$ between a
plaquette and the Wilson loop. (Right) Measurement of $E_{z}$ component of the
color flux \ in the y-z plane for the restricted $U(2)$ field. }%
\label{fig:measure}%
\end{figure}

We introduce a new formulation of the lattice YM theory of the minimal option,
which extracts the dominant mode of the quark confinement for $SU(3)$ YM
theory\cite{abeliandomSU(3),lattice2010}, since we consider the quark
confinement in the fundamental representation. Let $U_{x,\mu}=X_{x,\mu
}V_{x,\mu}$ be the decomposition of YM link variable, where $V_{x.\mu}$ could
be the dominant mode for quark confinement, and $X_{x,\mu}$ the remainder
part. The YM field and the decomposed new-variables are transformed by full
$SU(3)$ gauge transformation $\Omega_{x}$ such that $V_{x,\mu}$ is transformed
as the gauge link variable and $X_{x,\mu}$ as the site available:
\begin{subequations}
\label{eq:gaugeTransf}%
\begin{align}
U_{x,\mu}  &  \longrightarrow U_{x,\nu}^{\prime}=\Omega_{x}U_{x,\mu}%
\Omega_{x+\mu}^{\dag},\\
V_{x,\mu}  &  \longrightarrow V_{x,\nu}^{\prime}=\Omega_{x}V_{x,\mu}%
\Omega_{x+\mu}^{\dag},\text{ \ }X_{x,\mu}\longrightarrow X_{x,\nu}^{\prime
}=\Omega_{x}X_{x,\mu}\Omega_{x}^{\dag}.
\end{align}
The decomposition is given by solving the defining equation:
\end{subequations}
\begin{subequations}
\begin{align}
&  D_{\mu}^{\epsilon}[V]\mathbf{h}_{x}:=\frac{1}{\epsilon}\left[  V_{x,\mu
}\mathbf{h}_{x+\mu}-\mathbf{h}_{x}V_{x,\mu}\right]  =0,\label{eq:def1}\\
&  g_{x}:=e^{i2\pi q/N}\exp(-ia_{x}^{0}\mathbf{h}_{x}-i\sum\nolimits_{j=1}%
^{3}a_{x}^{(j)}\mathbf{u}_{x}^{(i)})=1, \label{eq:def2}%
\end{align}
where $\mathbf{h}_{x}$ is an introduced color field $\mathbf{h}_{x}%
=\xi(\lambda^{8}/2)\xi^{\dag}$ $\in\lbrack SU(3)/U(2)]$ with $\lambda^{8}$
being the Gell-Mann matrix and $\xi$ the $SU(3)$ gauge element. The variable
$g_{x}$ is undetermined parameter from Eq.(\ref{eq:def1}), $\mathbf{u}%
_{x}^{(j)}$ 's are $su(2)$-Lie algebra values, and $q_{x}$ an integer value
$\ 0,1,\cdots,N-1$. These defining equations can be solved exactly
\cite{exactdecomp}, and the solution is given by
\end{subequations}
\begin{subequations}
\label{eq:decomp}%
\begin{align}
X_{x,\mu}  &  =\widehat{L}_{x,\mu}^{\dag}\det(\widehat{L}_{x,\mu})^{1/N}%
g_{x}^{-1},\text{ \ \ \ }V_{x,\mu}=X_{x,\mu}^{\dag}U_{x,\mu}=g_{x}\widehat
{L}_{x,\mu}U_{x,\mu},\\
\widehat{L}_{x,\mu}  &  =\left(  L_{x,\mu}L_{x,\mu}^{\dag}\right)
^{-1/2}L_{x,\mu},\\
L_{x,\mu}  &  =\frac{N^{2}-2N+2}{N}\mathbf{1}+(N-2)\sqrt{\frac{2(N-1)}{N}%
}(\mathbf{h}_{x}+U_{x,\mu}\mathbf{h}_{x+\mu}U_{x,\mu}^{\dag})+4(N-1)\mathbf{h}%
_{x}U_{x,\mu}\mathbf{h}_{x+\mu}U_{x,\mu}^{\dag}\text{ .}%
\end{align}
Note that the above defining equations correspond to the continuum version:
$D_{\mu}[\mathcal{V}]\mathbf{h}(x)=0$ and $\mathrm{tr}(\mathbf{h}%
(x)\mathcal{X}_{\mu}(x))=0,$ respectively. In the naive continuum limit, we
have the corresponding decomposition $\mathbf{A}_{\mathbf{\mu}}(x)=\mathbf{V}%
_{\mu}(x)+\mathbf{X}_{\mu}(x)$ in the continuum theory\cite{SCGTKKS08} as
\end{subequations}
\begin{subequations}
\begin{align}
\mathbf{V}_{\mu}(x)  &  =\mathbf{A}_{\mathbf{\mu}}(x)-\frac{2(N-1)}{N}\left[
\mathbf{h}(x),\left[  \mathbf{h}(x),\mathbf{A}_{\mathbf{\mu}}(x)\right]
\right]  -ig^{-1}\frac{2(N-1)}{N}\left[  \partial_{\mu}\mathbf{h}%
(x),\mathbf{h}(x)\right]  ,\\
\mathbf{X}_{\mu}(x)  &  =\frac{2(N-1)}{N}\left[  \mathbf{h}(x),\left[
\mathbf{h}(x),\mathbf{A}_{\mathbf{\mu}}(x)\right]  \right]  +ig^{-1}%
\frac{2(N-1)}{N}\left[  \partial_{\mu}\mathbf{h}(x),\mathbf{h}(x)\right]  .
\end{align}

The decomposition is uniquely obtained as the solution of Eqs.(\ref{eq:decomp}%
), if color fields$\{\mathbf{h}_{x}\}$ are obtained. To determine the
configuration of color fields, we use the reduction condition which makes the
theory written by new variables ($X_{x,\mu}$,$V_{x,\mu}$) equipollent to the
original YM theory. Here, we use the reduction function
\end{subequations}
\begin{equation}
F_{\text{red}}[\mathbf{h}_{x}]=\sum_{x,\mu}\mathrm{tr}\left\{  (D_{\mu
}^{\epsilon}[U_{x,\mu}]\mathbf{h}_{x})^{\dag}(D_{\mu}^{\epsilon}[U_{x,\mu
}]\mathbf{h}_{x})\right\}  , \label{eq:reduction}%
\end{equation}
and color fields $\left\{  \mathbf{h}_{x}\right\}  $ are obtained by
minimizing the functional. It should be noticed that the gauge invariant
magnetic monopole $k_{\mu}$ is defined by using $V$-field:
\begin{subequations}
\begin{align}
\Theta_{\mu\nu}^{8}  &  :=-\arg\text{ \textrm{Tr}}\left[  \left(  \frac{1}%
{3}\mathbf{1}-\frac{2}{\sqrt{3}}\mathbf{h}_{x}\right)  V_{x,\mu}V_{x+\mu,\mu
}V_{x+\nu,\mu}^{\dag}V_{x,\nu}^{\dag}\right]  ,\\
k_{\mu}  &  =2\pi n_{\mu}:=\frac{1}{2}\epsilon_{\mu\nu\alpha\beta}%
\partial_{\nu}\Theta_{\alpha\beta}^{8},
\end{align}
which are derived from the non-Abelian Stokes' theorem\cite{KondoNAST}%
\cite{KondoShibata} and the Hodge decomposition of the field strength
$\mathcal{F}_{\mu\nu}[\mathbf{V}]$.$\ $It should be also noticed that, this
magnetic current (monopole) is non-Abelian magnetic monopole, which is defined
by $V$ field corresponding to the stability group $\tilde{H}=U(2)$.

Then, to investigate the color flux, we use the gauge invariant correlation
function. The color flux created by a quark-antiquark pair is measured by
using gauge invariant connected correlator of the Wilson loop \cite{Giacomo}:%
\end{subequations}
\begin{equation}
\rho_{W}:=\frac{\left\langle \mathrm{tr}\left(  U_{p}L^{\dag}WL\right)
\right\rangle }{\left\langle \mathrm{tr}\left(  W\right)  \right\rangle
}-\frac{1}{N}\frac{\left\langle \mathrm{tr}\left(  U_{p}\right)
\mathrm{tr}\left(  W\right)  \right\rangle }{\left\langle \mathrm{tr}\left(
W\right)  \right\rangle }, \label{eq:Op}%
\end{equation}
where $W$ represents a pair of quark and antiquark settled by Wilson loop in
Z-T plain, $U_{p}$ a plaquette variable as the probe operator to measure field
strength, and $L$ the Wilson line connecting the source $W$ and probe $U_{p}$,
and $N$ the number of color ($N=3$) (see the right panel\ of
Fig.\ref{fig:measure}). The symbol $\left\langle \mathcal{O}\right\rangle $
denotes the average of the operator $\mathcal{O}$ in the space and the
ensemble of the configurations. Note that this is sensitive to the field
strength rather than the disconnected one. Indeed, in the naive continuum
limit, the connected correlator $\rho_{W}$ is given by $\ \rho_{W}%
\overset{\varepsilon\rightarrow0}{\simeq}g\epsilon^{2}\left\langle
\mathcal{F}_{\mu\nu}\right\rangle _{q\bar{q}}:=\frac{\left\langle
\mathrm{tr}\left(  g\epsilon^{2}\mathcal{F}_{\mu\nu}L^{\dag}WL\right)
\right\rangle }{\left\langle \mathrm{tr}\left(  W\right)  \right\rangle
}+O(\epsilon^{4})$. Thus, the color filed strength is given by $\ F_{\mu\nu
}=\sqrt{\frac{\beta}{2N}}\rho_{W}$.

\section{Result}

We generate YM gauge configurations using the standard Wilson action on a
$24^{4}$ lattice with $\beta=6$.$2.$ The gauge link decomposition is obtained
by the formula given in the previous section, i.e., the color field
configuration is obtained by solving the reduction condition of minimizing the
functional eq(\ref{eq:reduction}) for each gauge configuration, and then the
decomposition is obtained by using the formula eq(\ref{eq:decomp}). In the
measurement of the Wilson loop, we apply the APE smearing technique to reduce
noises. To investigate the non-Abelian dual Meissner effect as the mechanism
of quark confinement, we measure correlators of the restricted $U(2)$ field,
$V_{x,\mu},$ as well as the original YM field. Note again that this restricted
$U(2)$-field and the non-Abelian magnetic monopole extracted from it reproduce
the string tension in the quark--antiquark potential \cite{lattice2010}%
\cite{abeliandomSU(3)}.

\begin{figure}[ptb]
\begin{center}
\vspace{-8mm} \includegraphics[
height=6.5cm,
angle=270
]
{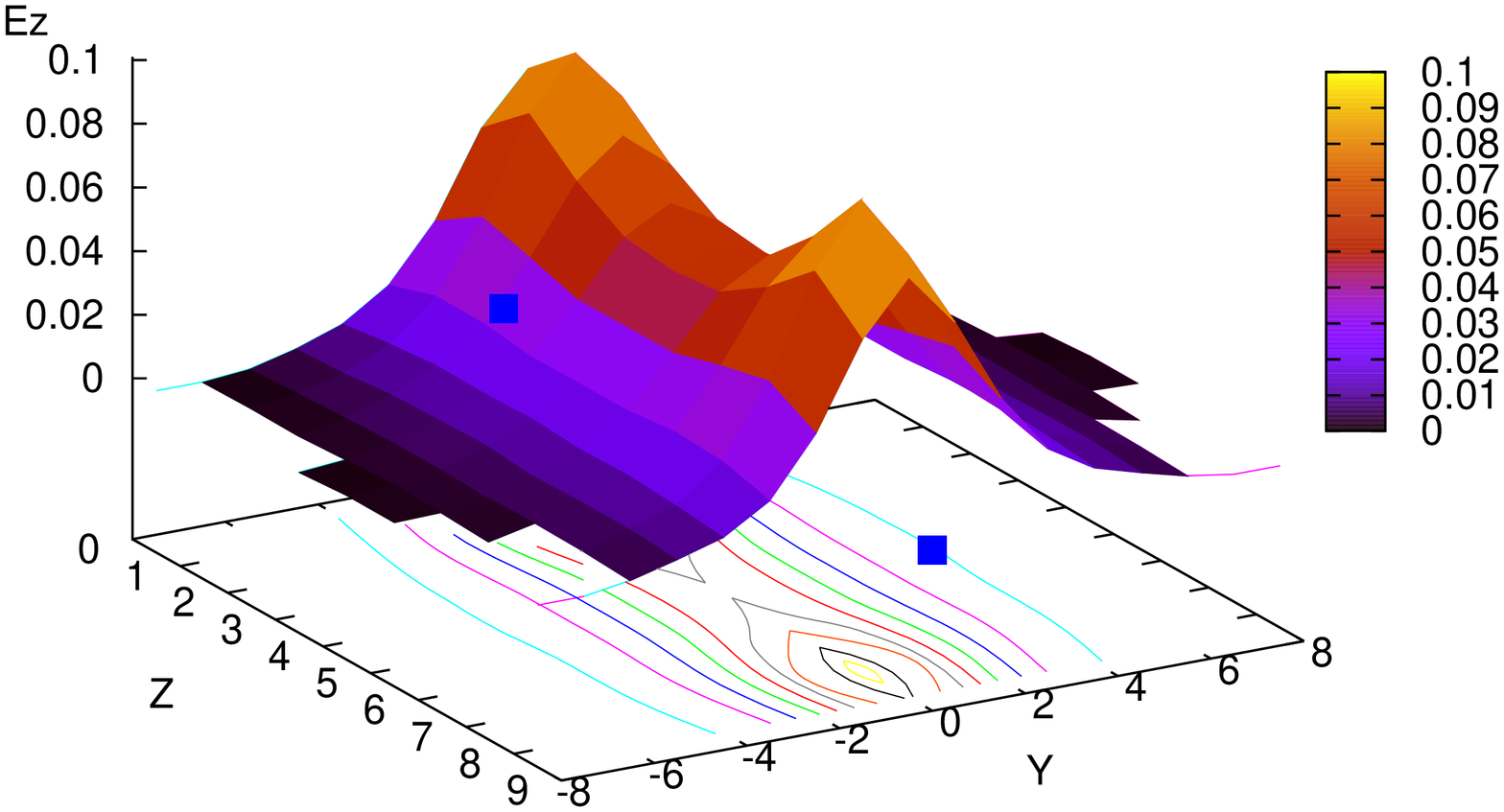} \ \includegraphics[
height=6.5cm,
angle=270
]
{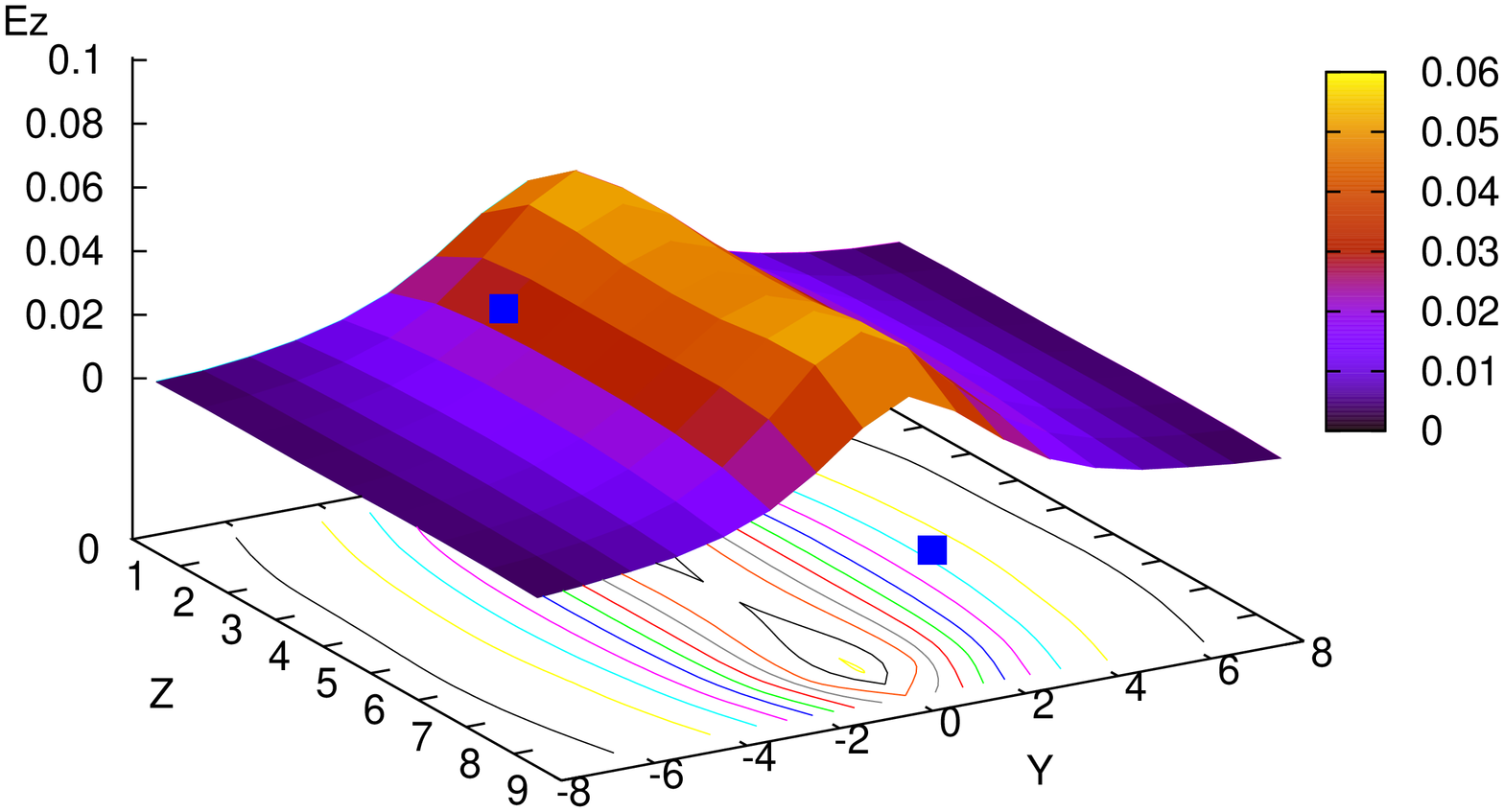} \vspace{-5mm}
\end{center}
\caption{The chromo-electric flux tube: (Left pannel) YM field, (Right pannel)
restricted $U(2)$ field. }%
\label{fig:fluxtube}%
\end{figure}

The right panel of Fig. \ref{fig:measure} shows the result of measurements of
the color flux obtained from the restricted $U(2)$-field, where the gauge link
variable $V_{\,x,\mu}$\ is used in eq(\ref{eq:Op}) instead of $U_{x,\mu}$.
Here the quark and antiquark source is introduce as $8\times8$ Wilson loop
($W$) in the Z-T plane, and the probe $(U_{p})$ is set at the center of Wilson
loop and moved along the Y-direction. The $E_{z}$ component only has non-zero
value as well as the original YM field, and it decreases quickly as away from
the Wilson loop. To know the shape of the color flux in detail, we explore the
distribution of color flux in the 2-dimensional plane. The right panel of
Fig.\ref{fig:fluxtube} shows the measurement of $E_{z}$ component of the color
flux, where the quark-antiquark source as $9\times11$ Wilson loop is placed in
the Z-T plane, and probe is displaced on the Y-Z plane at the midpoint of the
T-direction. \thinspace The chromo-electronic color flux is parallel to the
direction of quark and antiquark pair.

\begin{figure}[ptb]
\begin{center}
\vspace{-5mm}\includegraphics[
width=4.5cm
]
{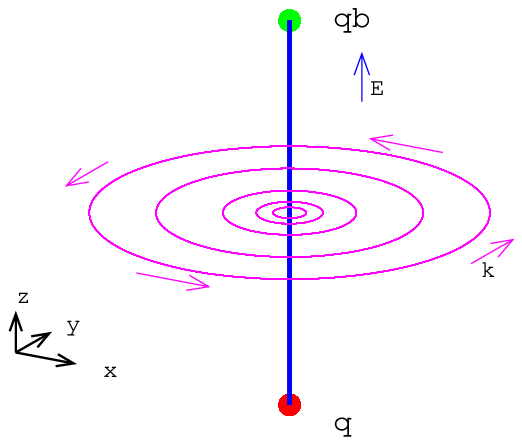} \ \ \includegraphics[
width=6.5cm
]
{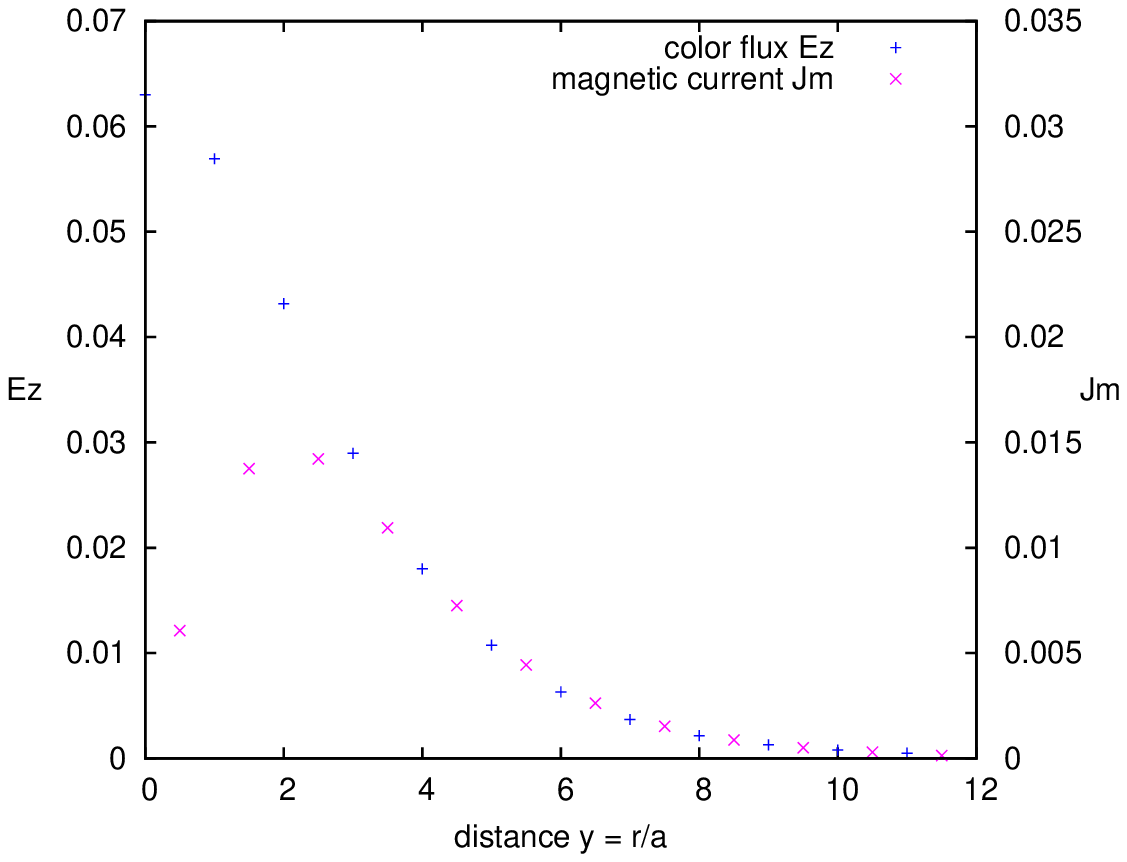} \vspace{-5mm}
\end{center}
\caption{{}The magnetic monopole (current) created by quark-antiquark pair.
(Left) the positional relationship between chromo-electric current and
magnetic current. (Right) The qualitative relation between chromo-electronic
current $E_{z}$ and magnetic current$~$\ $J_{m}=|\mathbf{k}|$ }%
\label{fig:Mcurrent}%
\end{figure}

Then, we investigate the relation between chromo-electronic flux and
\ magnetic current. From the Yang-Mills equation (Maxwell equation) for
$\mathbf{V}_{\mu}$ field, the magnetic monopole (current) can be calculated
as
\begin{equation}
\mathbf{k}={}^{\ast}dF[\mathbf{V}]\text{ ,}%
\end{equation}
where $F[\mathbf{V}]$ is the field strength 2-form of the $V_{\mu}$ field, $d$
the exterior derivative and $^{\ast}$ denotes the Hodge dual. Note that
non-zero magnetic current is a signal of the monopole condensation, be cause
of violation of the Bianchi identity. (Since the field strength is given by
the exterior derivative of $\mathbf{V}$ field (one-form), $F[\mathbf{V}%
]=d\mathbf{V}$, \ we obtain $\mathbf{k=}^{\ast}d^{2}\mathbf{V}$ $=0$).
Fig.\ref{fig:Mcurrent} shows the observed magnetic current in X-Y plain at the
midpoint of quark and anitquark pair in the Z-direction. The left panel shows
the positional relation between chromo-electric flux and \ magnetic current.
The right panel shows the magnitude of the magnetic chromo-electric flux (left
scale) and magnetic current (right scale). Therefore, we observed the monopole
condensation.

\begin{figure}[ptb]
\begin{center}
\vspace{-8mm}\includegraphics[
height=4.6cm
]
{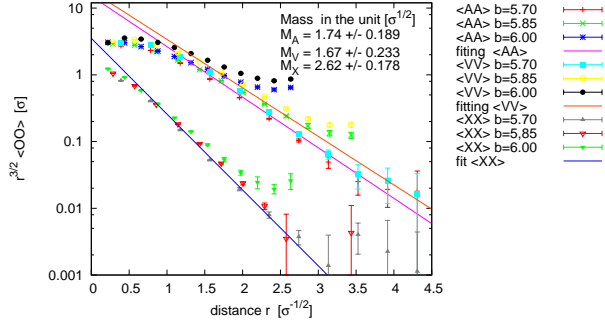} \vspace{-8mm}
\end{center}
\caption{The rescaled correration correlation functions $r^{3/2}\left\langle
O(r)O(0)\right\rangle ,$ $O=\mathbf{A,V,X}$ for $24^{4}$ lattice with
$\beta=5.7$, $5.85$, $6.0.$ The\ physical scale is in the unit of string
tenstion $\sigma_{\text{phys}}^{1/2}.$The correlation functions has profile od
cosh type becaus of the periodic boundary condition, and hence we use  data
withiin distance of the half size of lattice.}%
\label{fig.propagator}%
\end{figure}

Finally,\ we devote to study the 2-point correlation functions (propagators)
of the new variables and the original Yang-Mills field which are defined by
\begin{equation}
D_{OO}(x-y):=\left\langle O_{\mu}^{A}(x)O_{\mu}^{A}(y)\right\rangle \text{ for
}O_{\mu}^{A}(y^{\prime})\in\{\mathbf{V}_{x^{\prime},\mu},\mathbf{X}%
_{x^{\prime},\mu},\mathbf{A}_{x^{\prime},\mu}\},
\end{equation}
where an operator $O_{\mu}^{A}(x)$ is defined by linear type, e.g.,
$\mathbf{A}_{x^{\prime},\mu}:=(U_{x,\mu}-U_{x,\mu}^{\dag})/2g\epsilon$.
Fig.\ref{fig.propagator} shows the logarithmic plot of the scaled propagators
$r^{3/2}D_{OO}(r)$ with $r=|x-y|$ in the Landau gauge (LG), where the
correlation function is in the unit if string tension $\sigma_{\text{phys}}$,
and data of lattice spacing is from the TABLE I in Ref.\cite{Edward98}. The
correlation function $D_{VV}$ damps slowly and has almost the same damping as
$D_{AA}$, while the $D_{XX}$ damps quickly. Thus, from the view point of the
propagator, the $V$-field plays the dominant role in the deep IR region.

We estimate the mass of the propagator $D_{OO}(r)$ by using the massive
propagator in the Euclidian space, i.e., the Fourier transformation of behaves
for large $M_{X}$ as
\begin{equation}
D_{OO}(r)=\int\frac{d^{4}k}{(2\pi)^{4}}e^{ik(x-y)}\frac{3}{k^{2}+M_{O}^{2}%
}\simeq\frac{3\sqrt{M_{O}}}{2(2\pi)^{3/2}}\frac{e^{-M_{O}r}}{r^{3/2}}\text{
\ \ (}M_{o}r\gg1\text{)},
\end{equation}
and hence the scaled propagator $r^{3/2}D_{OO}(r)$ should be proportional to
$\exp(-M_{O}r).$ For parameter fitting of $M_{O}$ for $O=\{\mathbf{V}%
_{x^{\prime},\mu},\mathbf{A}_{x^{\prime},\mu}\}$, we use data in the region
$[2,4.5]$ and eliminated near the midpoint of the lattice to eliminate the
finite volume effect, and for $O=\mathbf{X}_{x^{\prime},\mu}$ we use the
region $[1,4].$ When we use $\sigma_{\text{phys}}=440MeV$, the preliminary
result shows
\begin{equation}
M_{A}\approx0.76GeV\text{, \ \ \ }M_{V}\simeq0.73GeV\text{, \ \ \ }M_{X}%
\simeq1.15GeV,
\end{equation}
which should be compared with result of the maximal case\cite{KKSSI07} in LLG,
and also result of the Abelian projection in the maximal Abelian
gauge\cite{Suganuma}.

\section{Summary and outlook}

We have studied the dual superconductivity for $SU(3)$ YM theory by using our
new formulation of YM theory on a lattice. We have extracted the restricted
$U(2)$ field from the YM field which plays a dominant role in confinement of
quark (fermion in the fundamental representation), i.e., the restricted $U(2)$
dominance and the non-Abelian magnetic monopole dominance in the string
tension, and measured the color flux for both the original YM field and the
restricted $U(2)$ field to confirm the non-Abelian dual superconductivity
picture. We have observed the dual Messier effect of $SU(3)$ YM theory, i.e.,
the flux tube, magnetic monopoles created by quark and antiquark pair. It is
important to determine the dual superconductivity, i.e., type I or type II,
which should be compared with border of type I and II\ to the SU(2) YM\ theory.

\subsection*{Acknowledgement}

This work is supported by Grant-in-Aid for Scientific Research (C) 24540252
from Japan Society for the Promotion Science (JSPS), and also in part by JSPS
Grant-in-Aid for Scientific Research (S) 22224003. The numerical calculations
are supported by the Large Scale Simulation Program No.T11-15 (FY2011) and
No.12-13 (FY2012) of High Energy Accelerator Research Organization (KEK).


\begin{thebibliography}{99}                                                                                               %


\bibitem {dualSC}Y. Nambu, Phys. Rev. D10, 4262(1974); G. 't Hooft, in: High
Energy Physics, edited by A.; Zichichi (Editorice Compositori, Bologna,
1975).; S. Mandelstam, Phys. Report 23, 245(1976).; A.M. Polyakov, Nucl. Phys.
B120, 429(1977).

\bibitem {Suzuki90}T. Suzuki and I. Yotsuyanagi, Phys. Rev. D42 4275 (1990)

\bibitem {stack94}J.D.Stack, S.D. Neiman and R.Wensley, Phys. Rev. D50 3399 (1994)

\bibitem {shiba}H. Shiba and T. Suzuki, Phys. Lett. B351 519 (1995)

\bibitem {greensite}J. Greensite, Prog. Part. Nucl. Phys. 51 1 (2003)

\bibitem {CFNS-C}Y.M. Cho, Phys. Rev. D 21, 1080 (1980). Phys. Rev. D 23, 2415
(1981); Y.S. Duan and M.L. Ge, Sinica Sci., 11, 1072(1979); L. Faddeev and
A.J. Niemi, Phys. Rev. Lett. 82, 1624 (1999); S.V. Shabanov, Phys. Lett. B
458, 322 (1999). Phys. Lett. B 463, 263 (1999).

\bibitem {KSM05}K.-I. Kondo, T. Murakami and T. Shinohara, Eur. Phys. J. C 42,
475 (2005); K.-I. Kondo, T. Murakami and T. Shinohara, Prog. Theor. Phys. 115,
201 (2006).

\bibitem {ref:NLCVsu2}S. Ito, S. Kato, K.-I. Kondo, A. Shibata, T. Shinohara,
Phys.Lett. B645 67-74 (2007)

\bibitem {ref:NLCVsu2-2}A. Shibata, S. Kato, K.-I. Kondo, T. Murakami, T.
Shinohara, S. Ito, Phys.Lett. B653 101-108 (2007)

\bibitem {kato:lattice2009}S. Kato, K-I. Kondo, A. Shibata and T. Shinohara,
PoS(LAT2009) 228.

\bibitem {SCGTKKS08}K.-I. Kondo, T. Shinohara and T. Murakami, Prog.Theor.
Phys. 120, 1 (2008)

\bibitem {lattce2007}Akihiro Shibata, Seiko Kato,Kei-Ichi Kondo, Toru
Shinohara \ and Shoichi Ito,

CHIBA-EP-166, KEK-PREPRINT-2007-50, POS(LATTICE2007) 331, arXiv:0710.3221 [hep-lat]

\bibitem {lattice2008}Akihiro Shibata, Seiko Kato,Kei-Ichi Kondo, Toru
Shinohara \ and Shoichi Ito, KEK-PREPRINT-2008-36, CHIBA-EP-173, 56 [hep-lat],
PoS(LATTICE 2008) 268

\bibitem {SCGTKKS08L}K.-I. Kondo, A.Shibata, T. Shinohara, T. Murakami, S.
Kato and S. Ito, Phys. Lett. B669, 107 (2008)

\bibitem {lattice2009}Akihiro Shibata, Kei-Ichi Kondo, Seikou Kato, Shoichi
Ito, Toru Shinohara, Nobuyui Fukui, PoS LAT2009 (2009) 232, arXiv:0911.4533 [hep-lat].

\bibitem {lattice2010}A. Shibata, K-I. Kondo,S. Kato and T. Shinohara,
PoS(Lattice 2010)286

\bibitem {exactdecomp}A. Shibata, K.-I. Kondo and T. Shinohara,
Phys.Lett.B691:91-98,2010, arXiv:0911.5294[hep-lat]

\bibitem {abeliandomSU(3)}Kei-Ichi Kondo, Akihiro Shibata, Toru Shinohara,
Seikou Kato, Phys.Rev. D83 (2011) 114016

\bibitem {KondoNAST}K.-I. Kondo, Phys.Rev.D77 085029 (2008)

\bibitem {KondoShibata}K.-I. Kondo and A. Shibata, CHIBA-EP-170,
KEK-PREPRINT-2007-73, arXiv:0801.4203[hep-th]

\bibitem {Cardaci2011}Mario Salvatore Cardaci, Paolo Cea, Leonardo Cosmai,
Rossella Falcone and Alessandro Papa, Phys.Rev.D83:014502,2011

\bibitem {Cardso}N. Cardoso, M. Cardoso, P. Bicudo, arXiv:1107.1355~[hep-lat]
(also lattice2011)

\bibitem {DualSC:KKSS2012}S. Kato, K.-I. Kondo, A. Shibata and T. Shinohara,
in preparation.

\bibitem {CeaCosmail2012}Cea, Cosmai and Papa, PRD86(054501) 2012

\bibitem {Giacomo}A. Di Giacomo, M. Maggiore, and S. Olejnik, Phys. Lett.
B236, 199 (1990); Nucl. Phys. B347, 441 (1990).

\bibitem {flusx:AP}Yoshimi Matsubara, Shinji Ejiri and Tsuneo Suzuki, NPB Poc.
suppl~34, 176~(1994)

\bibitem {Edward98}R.G.Edwards, U.M. Heller and T.R.Klassen, Phys. Rev. Lett.
80, 3448--3451 (1998)

\bibitem {KKSSI07}A.Shibata,S.Katob, K.-I. Kondoc,T. Murakamid,T.
Shinoharadand S. Ito, PoS(LATTICE 2007)331

\bibitem {Suganuma}Shinya Gongyo, Takumi Iritani, Hideo Suganuma, Phys.Rev.
D86 (2012) 094018
\end{thebibliography}
\end{document}